\documentclass[aps,twocolumn,amsmath,amssymb,preprintnumbers,longbibliography]{revtex4-1}
\usepackage{amsmath} \usepackage{amsfonts} \usepackage{amssymb}
\usepackage{bbm}
\usepackage{epsfig}
\usepackage{graphics}
\usepackage{graphicx}
\newcommand{\be}{\begin{equation}}
\newcommand{\ee}{\end{equation}}
\newcommand{\ba}{\begin{eqnarray}}
\newcommand{\ea}{\end{eqnarray}}
\newcommand{\nn}{\nonumber}

\begin{document}

\title[ ]{Modified gravity and coupled quintessence}

\author{C. Wetterich}
\affiliation{Institut  f\"ur Theoretische Physik\\
Universit\"at Heidelberg\\
Philosophenweg 16, D-69120 Heidelberg}

\begin{abstract}
The distinction between modified gravity and quintessence or dynamical dark energy is difficult. Many models of modified gravity are equivalent to models of coupled quintessence by virtue of variable transformations. This makes an observational differentiation between modified gravity and dark energy very hard. For example, the additional scalar degree of freedom in $f(R)$-gravity or non-local gravity can be interpreted as the cosmon of quintessence. Nevertheless, modified gravity can shed light on questions of interpretation, naturalness and simplicity. We present a simple model where gravity is modified by a field dependent Planck mass. It leads to a universe with a cold and slow beginning. This cosmology can be continued to the infinite past such that no big bang singularity occurs. All observables can be described equivalently in a hot big bang picture with inflation and early dark energy.
\end{abstract}

\maketitle

\section{Introduction}
\label{Introduction}

Einstein's equation 

\begin{equation}\label{1}
M^2(R_{\mu\nu}-\frac{1}{2}Rg_{\mu\nu})=T_{\mu\nu}
\end{equation}
expresses geometrical quantities on the left hand side in terms of matter and radiation on the right hand side. The basic geometrical quantity is the metric $g_{\mu\nu}$, with $R_{\mu\nu}$ and $R$ the Ricci tensor and curvature scalar formed from the metric and its derivatives. The energy momentum tensor $T_{\mu\nu}$ contains contributions from the particles of the standard model (``baryons'', neutrinos, radiation) and from dark matter.

The observation of the present accelerated expansion \cite{SN1,SN2} as well as indications for an inflationary epoch in very early cosmology tell us that equation \eqref{1} cannot be complete despite the numerous successful predictions of general relativity. One may supplement terms on the left or right side, as indicated by the dots
\begin{equation}\label{2}
R_{\mu\nu}-\frac{1}{2}Rg_{\mu\nu}+_{\dots}=\frac{1}{M^2}(T_{\mu\nu}+_{\dots}~).
\end{equation}
Additional contributions to the energy momentum tensor are usually called dark energy, whereas a change on the left hand side is associated with a modification of gravity or general relativity.

It is obvious that this distinction cannot be a particularly strict one since the validity of an equation does not depend on where one writes terms. The most prominent candidate for the explanation of an accelerated expansion, the cosmological constant, can be interpreted as an additional contribution to the energy momentum tensor $\Delta T_{\mu\nu}=\lambda g_{\mu\nu}$. This interpretation is suggested by the contribution of the effective potential of the Higgs scalar to $\lambda$, or similar for other scalar fields. We could write the cosmological constant term also on the left hand side and consider it as a modification of gravity - after all it influences the gravitational equations in ``empty space''. 

One may try a more concise definition of the meaning of modified gravity by requiring that the change of the Einstein tensor on the l.h.s. of eq. \eqref{1} involves derivatives of the metric, while terms with additional fields and no derivatives of $g_{\mu\nu}$ would contribute to $T_{\mu\nu}$. We will see, however, that modified gravity models defined in this way can often be rewritten in terms of different fields, frequently additional scalar fields. What appears in one field basis as a modification of gravity with terms involving derivatives of the metric shows up as dark energy with new fields and without metric derivatives in an other field basis. In particular, modified gravity theories that are consistent with the observed evolution of the universe are often equivalent to dynamical dark energy or quintessence. The borderline between modified gravity and dark energy becomes rather fuzzy. In fact, the first model of quintessence has originally been formulated as a modification of gravity \cite{CW3}.

The reason for this ambiguity between modified gravity and dark energy is connected to a basic property: observables depend on the dynamical degrees of freedom, but not on the choice of fields used to describe them (``field relativity''). For example, the metric may contain a scalar degree of freedom besides the graviton. This scalar is not distinguished from a ``fundamental scalar field'' (cosmon) which is the basic ingredient of quintessence.

These lecture notes will present several examples for the equivalence of modified gravity and quintessence. In particular, $f(R)$ gravity or a large class of non-local gravity models are equivalent to coupled quintessence \cite{CW2,ACQ}. We do not aim, however, to cover all possible modifications of gravity. More general modified gravity models may contain further non-scalar degrees of freedom (vectors of tensors), involve an infinite number of degrees of freedom, or give up the basic diffeomorphism symmetry underlying general relativity. 

Recent reviews of modified gravity can be found in refs. \cite{ReVO,Rev1,Rev2,Rev3,Rev3A,ReV4}. We concentrate here on the deep connection between modified gravity and coupled quintessence. This helps to understand many of the rich features of modified gravity in a simple and unified way. It also shows that many claims for observational distinguishability between modified gravity and quintessence are actually not justified. 

In sects. \ref{Basic setting} and \ref{Weyl scaling} we display our basic setting and discuss the field transformations that relate different versions of a given physical model. In sect. \ref{Brans-Dicke cosmology} we describe the cosmology of Brans-Dicke theory in the language of coupled quintessence. This points to strong observational bounds on the effective coupling $\beta$ between the cosmon and matter that will play an important role later. Sect. \ref{Scalar tensor models} discusses general scalar-tensor models with actions containing up to two derivatives. We highlight the importance of field-dependent particle masses in order to find models obeying the bounds on $\beta$. Sect. \ref{Slow Freeze Universe} discusses a simple three-parameter cosmological model along these lines which is compatible with all present observations from inflation to late dark energy domination. Formulated as a scalar-tensor theory (Jordan frame) it exhibits an unusual cosmic history. The universe shrinks during the 
radiation- and matter-dominated epochs and the evolution is always very slow. Cosmological solutions remain regular in the infinite past and there is no big bang singularity. On the other hand, the same model is characterized in the Einstein frame by a more usual big bang picture. This underlines that the field transformations that a crucial for these notes also incorporate important conceptual aspects.

In sect. \ref{Modified gravity with $f(R)$} we describe the equivalence of $f(R)$-modified gravity with coupled quintessence \cite{CNO,HS,KAS}. For constant particle masses the equivalent coupled quintessence models exhibit a large universal cosmon-matter coupling $\beta=1/\sqrt{6}$. This issue is a major problem for the construction of realistic $f(R)$ models. We sketch in sect. \ref{$f(R)$-gravity with varying particle masses} how a vanishing coupling $\beta=0$ can be obtained for $f(R)$-models with field-dependent particle masses. In sect. \ref{Non-local gravity} we turn to simple models of non-local gravity. Again, such models are equivalent to coupled quintessence. In sect. \ref{Higher derivative} we ask the general question to what extent modified gravity models which lead to second order field equations, as Horndeski's models \cite{Horn}, can find an equivalent description as coupled quintessence models. We find a huge class of such modified gravity models for which the scalar-gravity part is given by 
the action for quintessence, while additional information is 
contained in the details of the effective 
cosmon-matter coupling. Our conclusions are drawn in sect. \ref{Conclusions}. Parts of sects. \ref{Scalar tensor models}, \ref{Slow Freeze Universe} have overlap with work reported in refs. \cite{CWVG,Wetterich:2014eaa}.

\section{Basic setting}
\label{Basic setting}
We will assume that the theory which describes the late universe (say from radiation domination onwards) can be formulated as a quantum field theory. (This quantum field theory may be an effective theory embedded in a different framework as string theory.) We also restrict the discussion to the case where diffeomorphism symmetry (invariance under general coordinate transformations) is maintained. The most convenient way of specifying models is then the quantum effective action $\Gamma$ from which the field equations can be derived by variation. It is supposed to include all effects from quantum fluctuations. We can perform arbitrary changes of variables in $\Gamma$. They correspond to changes of variables in the differential field equations. All predictions of the model are contained in the field equations. A change of variables can therefore not affect any observable quantities. We will in the following heavily rely on this property of ``field relativity'' in order to demonstrate the equivalence of many 
modified gravity theories with coupled quintessence. (Note that on the level of the functional integral for a quantum theory a change of variables has two effects. It transforms the classical action and induces a Jacobian for the functional measure. The effective action is already the result of functional integration such that no Jacobian plays a role in the variable transformation.)

 We postulate that $\Gamma$ is invariant under general coordinate transformations and write it in the form 
\begin{equation}\label{3}
 \Gamma = \int d^4x\sqrt{g}({\cal L}_g+{\cal L}_m).
\end{equation}
Here ${\cal L}_g$ is the gravitational part, while the variation of $\sqrt{g}{\cal L}_m$ with respect to $g_{\mu\nu}$ yields the energy momentum tensor $T^{\mu\nu}$. Einstein's equation follows for 
\begin{equation}\label{4}
 {\cal L}_g=-\frac{M^2}{2}R,
\end{equation}
while ${\cal L}_m$ involves matter and radiation
\begin{equation}\label{5}
 {\cal L}_m={\cal L}_\text{standard model} +{\cal L}_\text{\rm dark matter}.
\end{equation}
Modified gravity corresponds to a more general form of ${\cal L}_g$. The simplest form of quintessence adds to ${\cal L}_m$ the contribution from a scalar field $\varphi(x)$, consisting of a potential $V(\varphi)$ and a kinetic term,
\begin{equation}\label{7}
 \triangle{\cal L}_m=\frac{1}{2}\partial^{\mu}\varphi\partial_{\mu} \varphi+V(\varphi).
\end{equation}
This scalar field is called the ``cosmon''. 

Simple modifications of gravity add to ${\cal L}_g$ terms involving higher powers of the curvature scalar as $R^2$. They can play an important role for inflation as in Starobinski's model \cite{Star}. Within higher dimensional theories the higher order curvature invariants have been employed for a mechanism of spontaneous compactification \cite{CWSC} and for a description of inflation as an effective transition from higher dimensions to four ``large'' dimensions \cite{SW1,SW2}. The field equations for actions where $R$ is replaced by an arbitrary function $f(R)$ have been investigated long ago \cite{Buch}. Modifications of gravity also arise if our four-dimensional world is a ``brane'' embedded in some higher-dimensional space \cite{DGP}. Higher-dimensional scenarios can be described in an equivalent four-dimensional setting, involving in principle infinitely many fields and in some cases non-local interactions. In the four-dimensional language typically both ${\
cal L}_g$ and ${\cal L}_m$ are modified 
simultaneously. We will concentrate in this lecture on simple four-dimensional models with only a few effective degrees of freedom. Many important aspects of modified gravity can be understood in this simple setting. We are mainly interested in the role of modified gravity for the present cosmological epoch and leave aside its potential relevance for the early inflationary epoch.

Modified gravity models have a long history. One of the most prominent historical models is Brans-Dicke theory \cite{BD}, where the reduced Planck mass $M$ in ${\cal L}_g$ is replaced by a scalar field $\chi(x)$. In this case both ${\cal L}_g$ and ${\cal L}_m$ get modified,
\ba\label{8}
{\cal L}_g&=&-\frac{\chi^2}{2}R,\\
 \triangle{\cal L}_m&=&\frac{1}{2}K\partial^{\mu} \chi \partial_{\mu}\chi.\label{9}
\ea
(Our choice of a scalar field $\chi$ differs from the original formulation in ref. \cite{BD}. The constant $K$ is related to the $\omega$-parameter in Brans-Dicke theory by $K=4\omega$.) Many aspects that are crucial for these notes can already be seen in Brans-Dicke theory, and we will discuss them in the next two sections. 

\section{Weyl scaling}
\label{Weyl scaling}

It is possible to express Brans-Dicke theory as a type of coupled quintessence model. For this purpose we perform a Weyl scaling \cite{Weyl,Di2} by using a different metric field $g'_{\mu\nu}$, related to $g_{\mu\nu}$ by 
\begin{equation}\label{19}
 g_{\mu\nu}=w^2g'_{\mu\nu}.
\end{equation}
Here the factor $w^2$ can be a function of other fields. Let us consider a scaling involving the scalar field $\chi$ without derivatives,  $w=w(\chi)$. The new curvature scalar $R'$ formed from $g'_{\mu\nu}$ and its derivatives is related to $R$ by 
\begin{equation}\label{20}
 R=w^{-2}\{R'-6(\ln w);^{\mu}(\ln w)_{;\mu}-6(\ln w); ^{\mu}{_{\mu}}\}.
\end{equation}
Here we denote by semicolons covariant derivatives, in particular
\ba\label{22}
(\ln w)_{;\mu}=\partial_\mu\ln w~,~
(\ln w)_;^\mu=g'^{\mu\nu}\partial_{\nu}\ln w.
\ea
The square root of the determinant of the metric, $g=-\det (g_{\mu\nu})$, transforms as 
\begin{equation}\label{21}
 \sqrt{g}=w^4\sqrt{g'}.
\end{equation}

We next make the specific choice 
\begin{equation}\label{24}
 w^2=\frac{M^2}{\chi^2},
\end{equation}
resulting in 
\be\label{25}
\sqrt{g}\chi^2R\rightarrow\sqrt{g'}M^2R'+ \text{ derivatives of } \chi.
\ee
The term ${\cal L}_g$ takes now the standard form \eqref{4} and the ``modification of gravity'' has been transformed away. As a counterpart, the kinetic term for $\chi$ is modified by replacing $\sqrt{g}\Delta {\cal L}_m\to \sqrt{g'}\Delta{\cal L}'_m$, 
\begin{equation}\label{30}
 \triangle{\cal L}'_m=\frac{M^2}{2}(K+6)\partial^{\mu}\ln \chi \phantom{.}\partial_{\mu}\ln\chi.
\end{equation}
For $K>-6$ the model describes gravity coupled to a scalar field. A canonical form of the scalar kinetic term $\Delta{\cal L}'_m=\partial^\mu\varphi\partial_\mu\varphi/2$ obtains for 
\be\label{30A}
\varphi=\sqrt{K+6}~M\ln\left(\frac{\chi}{M}\right).
\ee

The choice of the metric $g'_{\mu\nu}$ is called the Einstein frame. In the Einstein frame the Planck mass $M$ is a fixed constant that does not depend on any other fields. Cosmologies of two effective actions related by Weyl scaling are strictly equivalent, with all observables taking identical values \cite{CW1}. For a quantum field theory the concept of the quantum effective action $\Gamma$ is crucial for this statement. Its first functional derivatives, the field equations, describe exact relations between expectation values of quantum fields. Variable transformations as the Weyl scaling are transformations among these field values - they may be associated with ``field coordinate transformations''. Observables that can be expressed in terms of field values have to be transformed according to these variable transformations. For cosmology it is crucial that all quantities, including temperature $T$, particle masses $m$, or the coupling of particles to fields $\beta$, are transformed properly under Weyl 
scaling. It can then be established that suitable dimensionless ratios, as $T/m$, remain invariant under Weyl scaling \cite{CW1}. Dimensionless quantities are the only ones accessible to measurement and observation. One is therefore free to use the Einstein frame with metric $g'_{\mu\nu}$ or the ``Jordan frame'' \eqref{8} with metric $g_{\mu\nu}$ - both are equivalent, yielding the same results for dimensionless observable quantities. This has been verified by detailed studies of many observables \cite{CW1,DamE,FR3,Cat1,DeS,CWU}. We may summarize that physical observables cannot depend on the choice of fields used to describe them, a principle called ``field relativity'' \cite{CWU}. This principle extends to observables involving correlations, which can be found from higher functional derivatives of $\Gamma$. 

It is crucial that also the matter and radiation part ${\cal L}_m$ is transformed under Weyl scaling, due to the presence of the factor $\sqrt{g}$, or $g^{\mu\nu}$ in derivative terms. In general, not only the metric but also other fields appearing in ${\cal L}_m$ need to be transformed under Weyl scaling.  The electromagnetic gauge field $A_\mu$ needs no rescaling. Indeed the Maxwell kinetic term remains invariant since a factor $w^4$ from $\sqrt{g}$ cancels two factors $w^{-2}$ from the inverse metric $g^{\mu\nu}$ appearing in 
\be\label{30B}
{\cal L}_F=\frac14 F_{\mu\nu}F_{\rho\sigma}g^{\mu\rho}g^{\nu\sigma}.
\ee
For fermions, the factors of $w$ drop out of the kinetic term provided we combine the Weyl scaling \eqref{19} with a transformation of the fermion field
\begin{equation}\label{27}
 \psi=w^{-\frac{3}{2}}\psi'.
\end{equation}
This yields
\begin{equation}\label{28}
 \sqrt{g}\bar{\psi}\gamma^{\mu}\partial_{\mu}\psi\rightarrow\sqrt{g'}\bar{\psi}'\gamma^{\mu}\partial_{\mu}\psi'+...,
\end{equation}
where the dots denote a term containing a derivative of $\chi$, i.e. $\sqrt{g'}\bar\psi'\gamma^\mu\psi'\partial_\mu\chi$. For a model containing only massless gauge bosons and fermions the Weyl scaled version of Brans-Dicke theory describes standard gravity and a massless scalar field that has only derivative couplings. In this case $\varphi$ can be associated with the Goldstone boson of spontaneously broken dilatation or scale symmetry. 

For massive fermions the situation changes drastically. A mass term $m_F\sqrt{g}\bar\psi\psi$ transforms according to 
\begin{equation}\label{29}
m_F \sqrt{g}\bar{\psi}\psi\rightarrow
m'_F\sqrt{g'}\bar\psi'\psi'=m_F
\frac{M}{\chi}\sqrt{g'}\bar{\psi}'\psi'.
\end{equation}
We end with a non-derivative coupling of $\varphi$ to the fermion mass
\be\label{20A}
{\cal L}_{F,m}=m_F\exp 
\left(-\frac{\beta\varphi}{M}\right)
\bar\psi'\psi',
\ee
with cosmon-matter coupling\cite{CW1,CW2}
\be\label{29B}
\beta=\frac{1}{\sqrt{K+6}}=\frac{1}{\sqrt{4\omega+6}}.
\ee

\section{Brans-Dicke cosmology}
\label{Brans-Dicke cosmology}

For understanding the cosmological role of the coupling $\beta$ it is instructive to study the cosmology of the Brans-Dicke theory in the Einstein frame. We assume a homogeneous and isotropic Schwarzschild metric with scale factor $a(t),~H=\partial \ln a/\partial t$, and vanishing spatial curvature, coupled to a homogeneous scalar field~ $\varphi(t)$. The field equations for a fluid of massive particles read \cite{CW2,ACQ}
\ba\label{BD1}
&&H^2=\frac{1}{3M^2}(\rho+\frac12\dot{\varphi}^2),\\
&&\dot{\rho}+3H(\rho+p)+\frac\beta M(\rho-3p)\dot{\varphi}=0,\\\label{BD2}
&&\ddot{\varphi}+3H\dot{\varphi}=\frac{\beta}{M}(\rho-3p).\label{BD3}
\ea
For the radiation dominated epoch with $p=\rho/3$ the coupling $\beta$ plays no role. The field $\varphi$ settles rapidly to an arbitrary constant value and one finds standard cosmology. Additional massless fields for which $\beta$ vanishes do not change this situation.

Once particles become non-relativistic, however, and matter starts to dominate over radiation, the coupling $\beta$ leads to a modified cosmology. The field $\varphi$ evolves and particle masses change. After a transition period cosmology reaches a scaling solution which reads $(p=0)$
\be\label{BD4}
H=\frac{\eta}{t}~,~\dot{\varphi}=\frac{cM}{t}~,~\rho=\frac{f M^2}{t^2}.
\ee
Eqs. \eqref{BD1}-\eqref{BD3} become algebraic equations for $\eta,c$ and $f$, with solution
\ba\label{BD5}
\eta=\frac{2}{3+2\beta^2}~,~f=\frac{12-8\beta^2}{(3+2\beta^2)^2}~,~c=\frac{4\beta}{3+2\beta^2}.
\ea
This asymptotic solution exists for
\be\label{BD6}
\beta<\sqrt{\frac32}~~,~~\omega>-\frac43.
\ee

For $\beta$ of the order one one finds a scalar field dominated cosmology that is not compatible with observation. This becomes even more drastic for $\beta>\sqrt{3/2}$ where matter can be neglected as compared to the scalar kinetic energy. In contrast, for small $\beta$ the modification of the expansion remains small, with $\eta$ close to the standard value $2/3$. The most prominent cosmological effect concerns the time variation of the ratio of nucleon mass over Planck mass. Indeed, the field $\varphi$ has changed between matter-radiation equality and today by $\Delta\varphi=\varphi(t_0)-\varphi(t_{eq})$,
\be\label{BD7}
\Delta\varphi\approx 4\beta M\ln \left(\frac{t_0}{t_{eq}}\right),
\ee
with a corresponding change of the nucleon mass
\ba\label{BD8}
\frac{m_n(t_{eq})}{m_n(t_0)}\approx \exp 
\left(\frac\beta M\Delta\varphi\right)=
\left(\frac{t_0}{t_{eq}}\right)^{4\beta^2}
=z_{eq}{^{6\beta^2}}\approx(1100)^{\frac{3}{2\omega}}.\nn\\
\ea
The relative change of the nucleon mass $R_n\approx (3/2\omega)\ln(1100)$ bounds $\omega$ as a function of the observational bound $R_n<\bar R_n$ on the relative variation of the nucleon mass, 
\be\label{BD9}
\omega>\frac{10}{\bar R_n}\gtrsim 100 .
\ee

The upper bound on the relative variation of the nucleon mass $\bar R_n$ can be estimated from nucleosynthesis. (For Brans-Dicke theory no substantial change of the nucleon mass occurs between nucleosynthesis and matter radiation equality.) We evaluate
\be\label{VC1}
R_n=\frac{\Delta m_n}{m_n}=-\frac12
\frac{\Delta G_N}{G_N},
\ee
with $\Delta m_n=m_n(t_n)-m_n~,~m_n=m_n(t_0)$ and $t_n$ the time of nucleosynthesis. The second equation involves Newton's constant $G_N$. It reflects the fact that all particle masses vary $\sim m_n$ and only dimensionless ratios as $m^2_n G_N$ can influence the element abundancies produced during nucleosynthesis \cite{DSW}. We may use the bound from ref. \cite{DSW}
\be\label{VC2}
-0.19\leq \frac{\Delta G_N}{G_N}\leq 0.1
\ee
for a constraint $\bar R_n=0.1,\omega>100.$ This cosmological bound is weaker than the bound from solar system gravity experiments $\omega>10^4$ \cite{Ber}. On the other hand, this bound restricts the overall cosmological evolution. More precisely, the cosmological bound constrains a combination of $\beta$ and the change in the normalized cosmon field since nucleosynthesis,
\be\label{VC3}
-0.05\leq \frac\beta M\big(\varphi(t_n)-\varphi(t_0)\big)\leq 0.1.
\ee

\section{Scalar tensor models}
\label{Scalar tensor models}

The problem with $\varphi$-dependent particle masses in the Einstein frame persists for many scalar tensor models. There are two types of general solutions for this issue:
\begin{itemize}
\item [(i)] Particle masses in the Jordan frame are dependent on $\chi$ and scale $\sim\chi$. In the Einstein frame the particle masses are then independent of $\chi$ and $\beta$ vanishes \cite{CW3,CW1}.
\item[(ii)] The scalar field $\varphi$ changes very little, both in cosmology and locally. 
\end{itemize}
The simplest way to realize the second alternative is to add a potential $V(\chi)$ in the Jordan frame. After Weyl scaling one finds in the Einstein frame

\ba\label{ST1}
\sqrt{g}V(\chi)&=&\sqrt{g'}V'(\chi),\\
V'&=&w^4V=\frac{M^4}{\chi^4}V=\exp 
\left\{-\frac{4\varphi}{\sqrt{K+6}~M}\right\}V.\nn
\ea
If $V'(\varphi)$ has a minimum at $\varphi_0$ the cosmological solution will typically settle at this minimum at early time, such that there is no residual cosmic time variation of the ratio $m_n/M$. On the other hand, if $\varphi$ settles to $\varphi_0$ only after nucleosynthesis or continues evolving, the cosmological bound \eqref{VC3} has to be respected.

A local mass distribution acts as a source for the scalar field with strength $\beta/M$. This induces an additional scalar-mediated attraction. For a massless scalar field the relative strength of this interaction as compared to Newtonian gravity is $2\beta^2$. If the scalar mass
\be\label{ST2}
m_\varphi=\sqrt{\frac{\partial^2 V}{\partial\varphi^2}(\varphi_0)}
\ee
is smaller than the inverse size of the solar system the presence of this scalar interaction would be visible in post-Newtonian gravity experiments, limiting $\beta^2<2.5\cdot 10^{-5}$, cf. eq. \eqref{29B}. For larger $m_\varphi$ the additional exponential suppression of a Yukawa interaction allows for larger $\beta$. If $m_\varphi$ exceeds the inverse size of a massive object the scalar field $\varphi$ tends to settle inside the object at a value different from $\varphi_0$. Then the nucleon mass becomes density dependent, implying again upper bounds on $\beta$ \cite{DDC1}. For models predicting large $\beta$ and a small cosmological mass $m_\varphi$ there remains still the possibility that the local mass inside an object is substantially higher than the cosmological mass outside the object, due to non-linear effects. This is called chameleon effect \cite{DDC2}. We will see that many popular $f(R)$-theories lead to large $\beta$ and small $m_\varphi$.

In the remainder of this section we will concentrate on the alternative (i) with $\chi$-dependent particle masses. We will investigate a general class of scalar tensor theories with an effective action
\be\label{ST3}
\Gamma=\int_xg^{\frac12}
\left\{ -\frac12 F(\chi)R+\frac12 K(\chi)\partial^\mu\chi\partial_\mu\chi+V(\chi)\right\}.
\ee
This is the most general form for a scalar coupled to gravity which preserves diffeomorphism symmetry, provided that terms with four or more derivatives can be neglected. For a homogenous and isotropic Universe (and for vanishing spatial curvature) the field equations take the form \cite{CW1,CWVG}
\ba\label{VG6}
K(\ddot{\chi}&+&3H\dot{\chi})+\frac12\frac{\partial K}{\partial \chi}\dot{\chi}^2= 
-\frac{\partial V}{\partial \chi}+\frac12 \frac{\partial F}{\partial\chi}R
+q_\chi,\\
FR&=&F(12H^2+6\dot{H})\label{VG7}\\
&=&4V-\left(K+6\frac{\partial F}{\partial\chi^2}\right)\dot{\chi}^2\nn\\
&&-6\frac{\partial F}{\partial\chi^2}(\ddot{\chi}+3H\dot{\chi})\chi-12\frac{\partial^2 F}{(\partial\chi^2)^2}
\chi^2\dot{\chi}^2-T^\mu_\mu,\nn\\
&&\hspace{-1.1cm} F(R_{00}-\frac12 R g_{00})=3FH^2\label{VG8}\\
&&=V+\frac12K\dot{\chi}^2-6\frac{\partial F}{\partial\chi^2}H\chi\dot{\chi}+T_{00}.\nn
\ea

The r.h.s. of the field equations involves the energy-momentum tensor $T_{\mu\nu}$ and the incoherent contribution to the scalar field equation $q_\chi$. The general consistency relation between $q_\chi, T_{00}=\rho$ and $T_{ij}=p\delta_{ij}$ reads
\be\label{VG9}
\dot\rho+3H(\rho+p)+q_\chi\dot\chi=0.
\ee
For an ideal fluid of particles with a $\chi$-dependent mass $m_p(\chi)$ the explicit form of $q_\chi$ is given by
\be\label{VG10}
q_\chi=-\frac{\partial\ln m_p}{\partial\chi}(\rho-3p).
\ee
In particular, for $m_p(\chi)\sim\chi$ and $\rho-3p=m_p n_p$,
with $n_p$ the number density of particles, eq. \eqref{VG10} reads
\be\label{VG11}
q_\chi=-\frac{\rho-3p}{\chi}=-\frac{m_p}{\chi}n_p.
\ee

Let us consider the case where particle masses scale $m_p\sim \chi$ and concentrate on 
\be\label{ST4}
F(\chi)=\chi^2~,~K(\chi)=K.
\ee
A particular case is $V=\lambda\chi^4$. In this case the effective action \eqref{ST3} contains no parameter with dimension of mass or length. If, furthermore, all particle masses in ${\cal L}_m$ scale precisely $\sim\chi$ no mass scale appears in ${\cal L}_m$ either. Such models are scale invariant or dilatation invariant \cite{Fu,CW3}. Scale symmetry can be realized by a fixed point in the ``running'' of dimensionless couplings and mass ratios as a function of $\chi$. If the strong gauge coupling, normalized a momentum scale $q^2=\chi^2$, is independent of $\chi$, the ``confinement'' scale $\Lambda_{{\rm QCD}}$ scales $\sim\chi$. For a scale invariant potential for the Higgs doublet 
\begin{equation}\label{10}
 \mathcal{L}_h=\frac{\lambda_h}{2}(h^\dag h-\epsilon_h\chi^2)^2
\end{equation}
the minimum occurs for 
\begin{equation}\label{11}
 h_0\sim\chi,
\end{equation}
such that for constant Yukawa couplings one has 
\begin{equation}\label{12}
 m_e\sim\chi,
\end{equation}
and similar for quark and other charged lepton masses.

The cosmology of a model with exact scale symmetry is simple. After Weyl scaling the potential becomes $V'=\lambda M^4$ and particle masses are constant. The model describes a standard cosmology with cosmological constant $\lambda M^4$, coupled to an exactly massless Goldstone boson with derivative couplings, the dilaton. The dilaton settles to an arbitrary constant value in early cosmology and is not relevant for late cosmology \cite{CW3}. In particular, this type of model cannot account for dynamical dark energy.

The situation changes profoundly if we allow for violations of scale symmetry (dilatation anomaly) \cite{CW3}. For example, we may consider a cosmological constant in the Jordan frame, $V=V_0$, or a quadratic potential $V=\mu^2\chi^2$. In both cases the potential in the Einstein frame decays exponentially,
\be\label{ST5}
V=M^4\exp 
\left(-\frac{\alpha\varphi}{M}\right),
\ee
with $\alpha=4/\sqrt{K+6}$ for $V=V_0$ and $\alpha=2/\sqrt{K+6}$ for $V=\mu^2\chi^2$. (We absorb a multiplicative constant by a shift in $\varphi$.) The scalar ``cosmon'' field will roll down the potential, $\varphi(t\to\infty)\to\infty,~V(t\to\infty)\to 0$. Models of this type with constant particle masses in the Jordan frame lead to non-trivial cosmologies \cite{OB,Fo}. They are excluded, however, by the bounds on the time variation of $m_n/M$ since the coupling $\beta$ is large. 

At this point a simple setting for a realistic dynamical dark energy becomes visible. One may combine a dilatation anomaly in the potential, say $V=V_0$ or $V=\mu^2\chi^2$, with a scale invariant standard model of particle physics. If the charged lepton masses and quark masses as well as $\Lambda_{{\rm QCD}}$ all scale proportional to $\chi$, the nucleon and charged lepton masses as well as binding energies and cross sections become independent of $\varphi$ in the Einstein frame. All observational bounds on time varying fundamental couplings and apparent violations of the equivalence principle are obeyed. The first realistic model of dynamical dark energy or quintessence was actually a ``modified gravity'' of this type \cite{CW3}. Models of this type can also explain the recent increase in the fraction of dark energy $\Omega_h$ \cite{ABW,CWNEU}. Scale symmetry violation in the neutrino sector induced by a dilatation anomaly in the sector of heavy singlet fields entering by the seesaw mechanism can account 
for 
an 
increasing neutrino mass in the Einstein frame, $\beta<0$. This stops the evolution of $\varphi$ as soon as neutrinos become non-relativistic, typically around $z=5$. From this time on the cosmology looks very similar to a cosmological constant. 

Scalar tensor models that lead to dynamical dark energy for the present cosmological epoch \cite{CW3,CW1,EQ1,EQ2,EQ3,FA2,EQ4,EQ5} are sometimes called ``extended quintessence''. By virtue of Weyl scaling they are equivalent to a subclass of ``coupled quintessence'' \cite{CW2,ACQ,CQ2,CQ2a,CQ3,CQ4,CQ5,CQ6,CQ7}. Constant particle masses in the Jordan frame imply in the Einstein frame a universal coupling $\beta$ for all massive particles, while $\chi$-dependent masses offer more realistic perspectives. As compared to constant particle masses in extended quintessence, coupled quintessence is a more general concept where the cosmon-matter coupling can vary from one species to another. While the effective coupling $\beta_n$ to nucleons has to be very small, more sizeable couplings to dark matter are allowed $(\beta_{dm}\lesssim 0.1)$, and the cosmon-neutrino coupling can be large, say $\beta_\nu\approx 100$.

\section{Slow Freeze Universe}
\label{Slow Freeze Universe}

In this section we briefly describe a simple scalar-tensor model with only three cosmologically relevant dimensionless parameters \cite{Wetterich:2014eaa}. It is based on the effective action
\be\label{SF2}
\Gamma=\int d^4x\sqrt{g}
\left\{-\frac{\chi^2}{2}R+
\left(\frac{2}{\alpha^2}-3\right)
\partial^\mu \chi\partial_\mu\chi+V(\chi)\right\}.
\ee
The potential
\be\label{SF3}
V=\frac{\mu^2\chi^4}{m^2+\chi^2}~,~\lambda=\frac{\mu^2}{m^2}. 
\ee
shows a crossover between two scale invariant limits, one for $\chi\to 0$ with $V\approx\lambda\chi^4$, and the other for $\chi\to\infty$ with $V/\chi^4\to\mu^2/\chi^2\to 0$. The mass scales $\mu$ and $m$ violate scale symmetry. We take 
\be\label{SF1}
\mu=2\cdot 10^{-33}{\rm eV}
\ee
and $m\approx 10^6\mu$. The Planck mass $\chi$ being dynamical, no tiny dimensionless parameter for the cosmological constant appears in this model. 

For ``late cosmology'' after inflation we can approximate
\be\label{F1}
V=\mu^2\chi^2.
\ee
During radiation domination the universe shrinks \cite{CWU} according to a de Sitter solution with negative constant Hubble parameter
\be\label{SF16}
H=-\frac\alpha 2\mu.
\ee
In this period the value of the cosmon field $\chi$ increases exponentially according to 
\be\label{SF17}
\dot{s}=\frac{\dot\chi}{\chi}=\alpha\mu~,~\chi\sim\exp (\alpha\mu t).
\ee
Due to the shrinking of the universe with scale factor $a\sim 1/\sqrt{\chi}$ the energy density in radiation increases $\sim \chi^2$,
\be\label{SF18}
\rho_r=3\left(\frac{\alpha^2}{4}-1\right)\mu^2\chi^2,
\ee
similar to the potential and kinetic energy in the homogeneous scalar field which obey 
\be\label{SF19}
\rho_h=V+\frac{2}{\alpha^2}\dot\chi^2
=3\mu^2\chi^2.
\ee
This results in a constant fraction of early dark energy \cite{EDE,DR}
\be\label{SF20}
\frac{\rho_h}{\rho_r+\rho_h}=\Omega_e=\frac{4}{\alpha^2}.
\ee

While the temperature increases during radiation domination, $T\sim(\rho_r)^{\frac14}\sim \sqrt{\chi}$, the particle masses increase even faster $\sim\chi$. The equilibrium number density of a given species gets strongly Boltzmann-suppressed once a particle mass exceeds $T$. With Fermi scale $\langle h\rangle\sim\chi$ and $\Lambda_{QCD}\sim\chi$, as well as constant dimensionless couplings, the decay rates scale $\sim\chi$, and all cross sections and interaction rates scale with the power of $\chi$ corresponding to their dimension. As a consequence, nucleosynthesis proceeds as in usual cosmology, now triggered by nuclear binding energies and the neutron-proton mass difference exceeding the temperature as $\chi$ increases. The evolution of all dimensionless quantities is the same as in standard cosmology, once we measure time in units of the (decreasing) inverse nucleon mass. The resulting element abundancies are essentially the same as in standard cosmology. The only difference arises from the presence of a 
fraction 
of early dark energy \eqref{SF20}. This acts similarly to the presence of an additional radiation component, resulting in a lower bound on $\alpha$ from nucleosynthesis \cite{CW3,CW2,BS2,BHM}. Later on, protons and electrons combine to hydrogen once the atomic binding energy (increasing $\sim\chi$) exceeds the temperature $T\sim\sqrt{\chi}$. Up to small effects of early dark energy the quantitative properties of the CMB-emission are the same as in standard cosmology. The effect of early dark energy on the detailed distribution of CMB-anisotropies gives so far the strongest bound on $\alpha,\alpha \gtrsim 10$, \cite{A2a,A2b,Re,Sievers:2013wk,A2d,PL}. 

The ratio of matter to radiation energy density increases as $\rho_m/\rho_r\sim\chi a$, with $a\sim\chi^{-\frac12}$ during radiation domination $(Ta=const.)$. This triggers the transition to a matter dominated scaling solution once $\rho_m$ exceeds $\rho_r$, given again by a shrinking de Sitter universe
\be\label{SF21}
H=-\frac{\alpha\mu}{3\sqrt{2}}~,
~\dot{s}=\frac{\alpha\mu}{\sqrt{2}}~,
~\rho_m=\frac23(\alpha^2-3)\mu^2\chi^2,
\ee
with a constant fraction of early dark energy $\Omega_e=3/\alpha^2$. Observations of redshifts of distant galaxies are explained by the size of atoms shrinking faster than the distance between galaxies \cite{CWU,Na1,Na2,Na3}, resulting in an increase of the relevant ratio $\sim a\chi$. 

The transition to the present dark energy dominated epoch can be triggered by neutrinos. Assume that the heavy singlet scale entering the neutrino masses by the seesaw mechanism decreases with increasing $\chi$. Neutrino masses will then grow faster than $\chi$, with positive
\be\label{55A}
\tilde\gamma(\chi)=\frac12
\frac{\partial\ln\big(m_\nu(\chi)/\chi\big)}{\partial\ln \chi}.
\ee
The value of $\tilde \gamma$ in the present epoch will be the third dimensionless cosmological parameter of our model besides $\alpha$ and $\mu/m$. Together with the present neutrino mass it determines the present dark energy density. 

In a rather recent cosmological epoch $(z\approx 5)$ the neutrinos become non-relativistic. For $\tilde\gamma\gg 1$ the increase of their mass faster than $\chi$ stops effectively the time evolution of the cosmon field. The dark energy density $\rho_h$ remains frozen at the value it had at this moment, relating it to the average neutrino mass. More precisely, the cosmological solution oscillates around a very slowly evolving ``average solution'' for which the r.h.s. of eq. \eqref{VG6} vanishes to a good approximation, $V=\tilde \gamma\rho_\nu$. This yields for the homogeneous dark energy density $\rho_h$ the interesting quantitative relation \cite{ABW}
\be\label{SF22}
\rho_h^{\frac14}=1.27\left(\frac{\tilde \gamma m_\nu}{{\rm eV}}\right)^{\frac14}10^{-3}{\rm eV}.
\ee
(Present neutrino masses on earth may deviate from the value of $m_\nu$ according to the cosmological average solution, due to oscillations and a reduction factor for neutrinos inside large neutrino lumps \cite{GNQ10,GNQ5A}. Cosmological bounds on $m_\nu$ are modified due to the mass variation.) 

For low redshift $z\lesssim 5$ cosmology is very similar to the $\Lambda$CDM-model with an effective equation of state for dark energy (more precisely the coupled cosmon-neutrino fluid) very close to $-1$,
\be\label{2SF23}
w=-1+\frac{\Omega_\nu}{\Omega_h}=-1+\frac{m_\nu(t_0)}{12{\rm eV}}.
\ee
An important observational distinction to the $\Lambda$CDM-model is the clumping of the neutrino background on very larges scales which may render it observable \cite{GNQ2,GNQ8,GNQ9,GNQ10}. The parameter $\mu$ in eq. \eqref{SF1} obtains from the observed value of the present dark energy density $\sqrt{\rho_h}=(2\cdot 10^{-3}$eV$)^2\approx\sqrt{V}=\mu M$. This also fixes $\tilde\gamma m_\nu=6.15$eV.

Primordial cosmology corresponds to an inflationary epoch. Matter and radiation play no role and we solve the field equations \eqref{VG6}-\eqref{VG8} with $T_{\mu\nu}=0,q_\chi=0$. One finds a scaling solution without a big bang singularity that can be continued to $t\to-\infty$,
\be\label{F2}
\chi=\left(\frac{-2m^3}{\sqrt{3}\alpha^2\mu t}\right)^{\frac13}~,~
H=\left(\frac{-2\mu^2}{9\alpha^2 t}\right)^{\frac13}~,~
\frac{\dot{\chi}}{\chi}=-\frac{1}{3t}.
\ee
The spectrum of primordial density fluctuations generated during inflation will be discussed below. 

A universe shrinking during radiation and matter domination was much colder in the past than the present background radiation. Its shrinking was very slow, with $|H|\approx \alpha\mu$ only slightly faster than the present expansion rate. During inflation the expansion was even slower, cf. eq. \eqref{F2}. The typical time scale of the universe was never much shorter than $10^{10}$yr. Despite the unusual aspects of such a ``slow freeze'' picture of the evolution of the universe no present observation is in contradiction to it.

For a quantitative discussion of observables it is useful to perform a Weyl scaling in order to bring this model to the form \eqref{3}-\eqref{7}. In the Einstein frame the potential decays exponentially for large $\varphi$
\be\label{F6}
V'=\lambda M^4\left[1+\exp\left(\frac{\alpha\varphi}{M}\right)\right]^{-1}.
\ee
Particle masses except for the neutrinos do not depend on $\varphi$, while the cosmon-neutrino coupling 
\be\label{F7}
\beta=-M\frac{\partial\ln m_\nu}{\partial\varphi}=-\frac{\tilde\gamma}{\alpha}
\ee
realizes growing neutrino quintessence. 

A quantitative discussion of the spectrum of density fluctuations is straightforward in the Einstein frame. For the inflationary epoch, our model can be treated in the slow roll approximation. For fluctuations corresponding to the present scale of galaxies or clusters, which have crossed the horizon $N$ $e$-foldings before the end of inflation, one finds for the spectral index $n$
\be\label{F3}
n=\frac{1}{2N}=0.96-0.967,
\ee
while the tensor amplitude $r$ is very small 
\be\label{F4}
r=\frac{8}{N^2\alpha^2}<3\cdot 10^{-5}.
\ee
A realistic amplitude for the primordial density fluctuations is found for 
\be\label{F5}
\frac{\mu}{m}=\frac{5}{N\alpha}\cdot 10^{-4}.
\ee
The spectrum of primordial density fluctuations of our model is compatible with Planck-results \cite{PL}.

Our model has no more free parameters than the $\Lambda$CDM-model and is therefore subject to many observational tests. Its compatibility with all present observations demonstrates how a simple modification of gravity can lead to a rather natural setting with a unified description of inflation and present dark energy. The naturalness of the simple quadratic potential for large $\chi,V=\mu^2\chi^2$, may look less obvious if the model would be originally formulated in the Einstein frame with a potential \eqref{F6}. While we could add a cosmological constant to $V(\chi)$ without affecting the late time behavior for large $\chi$, an addition of a constant to eq. \eqref{F6} would drastically change the late time cosmology. Thus the issue of naturalness of an asymptotically vanishing cosmological constant looks very different in modified gravity (Jordan frame) or the associated standard gravity (Einstein frame).

\section{Modified gravity with $f(R)$}
\label{Modified gravity with $f(R)$}

Let us next discuss $f(R)$-theories, where ${\cal L}_g$ takes the form 
\be\label{F1a}
{\cal L}_g=  -\frac{M^4}{2} f(y)~,~y=\frac{R}{M^2}.
\ee
We will see that they are equivalent to models of coupled quintessence with a coupling $\beta=1/\sqrt{6}$. Due to their rather simple structure they are among the most popular models of modified gravity \cite{Buch,Hw,FR4,FR5,FR6,FA1,FR7,FR8,FR9,AFR1}.

We start with a simple example where $f$ contains terms linear and quadratic in $R$,
\ba\label{F2a}
 \Gamma[g_{\mu\nu}]=\int_x\sqrt{g}
\left\{-\frac{cM^2}{2}R-\frac{\alpha}{2}R^2\right\}~,~f(y)=cy+\alpha y^2.\nn\\
\ea
This includes the model used by A. Starobinski \cite{Star} in his early discussion of the inflationary universe. It is straightforward to see that this model is equivalent to a scalar model with 
\ba\label{39}
\Gamma[\phi,g_{\mu\nu}]=\int_x\sqrt{g}
\left\{-\frac{cM^2}{2}R-\frac{\alpha}{2}R^2+\frac{\alpha}{2}
\left(\frac{\phi}{\alpha}-R\right)^2\right\}.\nn\\
\ea
Indeed, the scalar field equation,
\be\label{40}
 \frac{\delta\Gamma}{\delta \phi}=0,
\ee
has a general solution
\be\label{41}
\phi=\alpha R.
\ee
Reinsertion into the effective action yields eq. \eqref{F2a}. Expanding the last term in eq. \eqref{39} yields the equivalent scalar-gravity model
\ba\label{46}
 \Gamma[\phi,g_{\mu\nu}]=\int_x\sqrt{g}
\left\{V(\phi)-\frac{M^2}{2}
\left(c+\frac{2\phi}{M^2}\right)R\right\},
\ea
with potential
\be\label{47}
 V(\phi)=\frac{1}{2\alpha}\phi^2.
\ee
At this stage the modified gravity model \eqref{F2a} has been transformed into a scalar-tensor model \eqref{46}.

We next perform a Weyl scaling with 
\be\label{48}
 w^2=\frac{1}{c+\frac{2\phi}{M^2}},
\ee
resulting in 
\ba\label{49}
&&\Gamma[\phi',g'_{\mu\nu}]=\int_x\sqrt{g'}
\left\{V'\right.\nn\\
&&\qquad~~\left.-\frac{M^2}{2}
\left(R'-\frac32(\ln w^2);{^\mu}(\ln w^2);_\mu\right)\right\},
\ea
with 
\be\label{50}
V'=w^4V=\frac{\phi^2}{2\alpha\left(c+\frac{2\phi}{M^2}\right)^2}.
\ee
The canonical normalization of the scalar kinetic term obtains for 
\be\label{51}
 \varphi=\sqrt{\frac{3}{2}}M\ln
\left( c+\frac{2\phi}{M^2}  \right),
\ee
corresponding to 
\be\label{36}
 w^2=exp\left\{-\sqrt{\frac{2}{3}}\frac{\varphi}{M}\right\}.
\ee
The modified gravity model appears now as a model of quintessence without any modification of gravity,
\be\label{52}
 \Gamma[\varphi,g'_{\mu\nu}]=\int_x\sqrt{g'}
\left\{V'+\frac12\partial^\mu\varphi\partial_\mu\varphi-\frac{M^2}{2}R'\right\}.
\ee
The potential decays exponentially for large $\varphi$
\be\label{53}
 V'(\varphi)=\frac{M^4}{8\alpha}
\left(1-c\exp 
\left(-\sqrt{\frac23}\frac{\varphi}{M}\right)\right)^2.
\ee
We take $\alpha>0$ such that the potential is bounded from below. 

It is instructive to expand the potential for small $\varphi$
\ba\label{54}
&&V'(\varphi)=\frac{M^4}{8\alpha}\\
&&\left\{(1-c)^2+\sqrt{\frac83}c(1-c)
\frac{\varphi}{M}+\frac23 c(2c-1)\frac{\varphi^2}{M^2}+\dots
\right\}.\nn
\ea
For $c=1$ the leading term is the quadratic
\be\label{55}
 V'(\varphi)=\frac{M^4}{12\alpha}\varphi^2+\dots
\ee
with scalar mass given by 
\be\label{56}
m_\varphi=\frac{M}{\sqrt{6\alpha}}. 
\ee
For $\alpha$ of the order one this mass turns out to be of the order of the Planck mass. In this case the scalar field settles very early in cosmology to the minimum of the potential at $\varphi=0$. Subsequently, the potential $V'$ plays no role for late cosmology. Cosmology is described by standard gravity coupled to a massive scalar field. The situation is similar for the corresponding modification of gravity. The term $\sim \alpha R^2$ in eq. \eqref{F2a} can play a role during inflation \cite{Star}, but is irrelevant for late cosmology. If one wants to have the term $\sim\alpha R^2$ to play a role in the present cosmological epoch one needs a huge value of $\alpha$ such that $\alpha R$ becomes comparable to $M^2$,
\be\label{F3a}
\alpha\approx 10^{60}.
\ee

This points to a very general issue for $f(R)$-theories: The deviations from Einstein's equation play a role in present cosmology only if the expansion in derivatives involves huge coefficients or diverges. In other words, any function $f(y)$ which admits a Taylor expansion around $f(y)$ with coefficients that are substantially smaller than $10^{60}$ leads to modifications of gravity that are not observable in the present cosmological evolution. This remark extends to more general effective actions, involving, for example, $R_{\mu\nu}R^{\mu\nu}$. 

For $c>0$ the potential has a minimum for a finite value of $\varphi$
\be\label{FRA}
\varphi_{\rm min}=\sqrt{\frac32}M\ln c.
\ee
We observe that at the minimum the effective cosmological constant vanishes
\be\label{FKB}
V'(\varphi_{\rm min})=0.
\ee
The scalar mass \eqref{56} is independent of $c$. For $c<0$ the minimum of $V'$ occurs for $\varphi\to \infty$, with 
\be\label{FRC}
V(\varphi\to\infty)=\frac{M^4}{8\alpha}.
\ee
In this case the scalar mass vanishes in the asymptotic limit. A realistic effective cosmological constant would require
\be\label{FRD}
\alpha\approx 10^{120}.
\ee

A major problem for $f(R)$-models is the universal large coupling $\beta=1/\sqrt{6}$ of the cosmon to all massive particles in the Einstein frame. Indeed, the Weyl scaling will take for all $f(R)$-models the form \eqref{36}. This implies for the nucleon mass in the Einstein frame
\be\label{57}
 m'_n=w m_n=\exp 
\left\{-\frac{1}{\sqrt{6}}\frac{\varphi}{M}\right\}m_n,
\ee
resulting in a cosmon-nucleon coupling
\be\label{58}
 \beta_n=-M\frac{\partial}{\partial\varphi}\ln m'_n=\frac{1}{\sqrt{6}}.
\ee
Thus $f(R)$-theories are equivalent to coupled quintessence. In order to obey the observational bound \eqref{VC3} on $m_n/M$ the cosmon is allowed to vary only by a tiny amount since nucleosynthesis. Furthermore, unless the cosmon mass is large enough, the large value $\beta_n=1/\sqrt{6}$ contradicts post-Newtonian gravity measurements in the solar system. The cosmological scalar mass is typically very small, however, if the modifications of gravity are important in present cosmology (e.g. eq. \eqref{56} with huge $\alpha$). Due to this clash,realistic models need to invoke the chameleon mechanism \cite{DDC2}. The combination of the absence of a Taylor expansion (with moderate coefficients) and the need for the chameleon mechanism limits severely the choice of realistic functions $f(y)$. At the end, realistic functions are very close to $f(y)=c_0+y$, with $c_0=\lambda/M^4$ related to the cosmological constant $\lambda$. In the next section we will sketch how part of these problems can be avoided for $f(R)$ 
theories with field dependent particle masses. 

We end this section by a short discussion of the general map from an $f(R)$-theory to coupled quintessence. Consider a scalar-tensor theory with 
\be\label{GF1}
\Gamma=\int_x\sqrt{g}\{-\phi R+V(\phi)\}.
\ee
The solution of the field equation for the scalar field expresses $\phi(R)$ as a function of $R$, given implicitly by
\be\label{GF2}
\frac{\partial V}{\partial\phi}=R.
\ee
For $\partial^2 V/\partial\phi^2\neq 0$ this solution is unique. Insertion of $\phi(R)$ into the action \eqref{GF1} yields an equivalent $f(R)$-theory \eqref{F1a} with 
\be\label{GF3}
f\left(\frac{R}{M^2}\right)=\frac{2}{M^4}\big\{R\phi(R)-V\big(\phi(R)\big)\big\}.
\ee

By virtue of eq. \eqref{GF2} the function $f(y)=f(R/M^2)$obeys the relation
\be\label{GF4}
\frac{\partial f(y)}{\partial y}=\frac{2\phi(R)}{M^2}.
\ee
The construction above associates to a given potential $V(\phi)$ an equivalent $f(R)$-model. Inversely, for a given $f(y)$ eqs. \eqref{GF3}, \eqref{GF4} yield the potential $V(\phi)$ as a Legendre transform
\be\label{GF5}
V(\phi)=\frac{M^4}{2}
\left(y\frac{\partial f(y)}{\partial y}-f(y)\right),
\ee
with $y(\phi)$ given by eq. \eqref{GF4}. This holds provided eq. \eqref{GF4} has a unique solution, i.e. for $\partial^2 f/\partial y^2\neq 0$. 

A Weyl scaling brings finally the action \eqref{GF1} to the standard form \eqref{52}. Due to the absence of a kinetic term in eq. \eqref{GF1} the dependence of the conformal factor $w$ on the normalized scalar field $\varphi$ is universal,
\be\label{GF6}
w^2=\frac{M^2}{2\phi}=\exp 
\left\{-\sqrt{\frac23}\frac\varphi M\right\}.
\ee
As a consequence, $f(R)$-theories with constant particle masses are found to be equivalent to coupled quintessence, with a universal coupling $\beta=1\sqrt{6}$ given by eq. \eqref{58}. For the normalized scalar field in the Einstein frame the potential is related to $f(y)$ by 
\be\label{37}
 V'(\varphi)=\frac{M^2}{2} \phantom{x} \frac{Rf'-f}{(f')^2}.
\ee

As an example, we may consider 
\be\label{EX1}
 f(y)=f_0y^{\gamma}.
\ee
Eq. \eqref{GF4} implies
\be\label{EX2}
\phi=\frac{\gamma f_0M^2}{2}
\left(\frac{R}{M^2}\right)^{\gamma-1}~,~
R=M^2
\left(\frac{2\phi}{\gamma f_0M^2}\right)^{\frac{1}{\gamma-1}},
\ee
and the potential in the scalar-tensor model reads
\be\label{EX3}
 V(\phi) = \frac{M^4(\gamma-1)}{2} f(y) = \frac{M^4 (\gamma-1) f_0}{2} \left(\frac{2\phi}{\gamma f_0 M^2}\right)^\frac{\gamma}{\gamma-1}.
\ee
Weyl scaling leads in the Einstein frame to an additional factor $(M^2/2\phi)^2$ for $V'$, such that 

\be\label{EX4}
 V'=\frac{M^4(\gamma-1)}{2\gamma}(\gamma f_0)^{-\frac{1}{\gamma-1}} \left(\frac{M^2}{2\phi}\right)^{1-\frac{1}{\gamma-1}}.
\ee

For the particular ``critical'' value $\gamma=2$ the potential $V'$ is constant. For $1<\gamma<2$ the minimum of $V'$ occurs for $\phi=0$, $V'(\phi=0)=0$. On the other hand, for $\gamma>2$ the potential takes its minimal value for $\phi\rightarrow\infty$, with

\be\label{EX5}
V'(\phi\rightarrow\infty)=0.
\ee
With
\be\label{EX6}
\phi=\frac{M^2}{2}\exp\left\{ \sqrt{\frac{2}{3}} \frac{\varphi}{M}\right\}
\ee
the limit $\phi\rightarrow\infty$ corresponds to $\varphi\rightarrow\infty$ and we observe that the potential $V'(\varphi)$ decays to zero exponentially. These models are of the same type as the one discussed in sect. VI, using in \eqref{SF2} the identifications $\phi=2\chi^2$, $\alpha^2=2/3$, and $V(\chi)=V'(\phi=2\chi^2)$. 

We observe that the addition of a cosmological constant $\bar\lambda_c$ in the effective action for modified gravity results in 
\be\label{EX7}
f(y)=f_0 y^{\alpha}-e_0, \phantom{X} \bar{\lambda}_c=\frac{e_0M^4}{2}.
\ee
After Weyl scaling this adds to $V'$ a part 

\be\label{EX8}
\Delta V'=\frac{e_0M^8}{8\phi^2} .
\ee
This becomes irrelevant for large $\phi$. Modified gravity theories with $\gamma>2$ are an example for a self-tuning of the cosmological constant to zero as a consequence of the asymptotic cosmological solution for large time.

For $\gamma=1$ one has Einstein gravity without an additional scalar degree of freedom. For $0<\gamma<1$ and $f_0>0$ the potential $V'$ is negative, diverging for $\phi\rightarrow0$. For negative $f_0$ one finds negative $\phi$ such that the gravitational constant would have a wrong sign, leading to instability. The range $0<\gamma<1$ does not seem to lead to a reasonable cosmology. We may, however, consider the values $\gamma<0$, $f_0<0$, with positive $\gamma f_0$ and $\phi$. The potential $V'$ is now again positive, decaying to zero for $\phi\rightarrow\infty$. The behaviour is similar as for $\gamma>2$ and $f_0>0$. We conclude that $f(R)$-models could lead to interesting cosmologies with a dynamical self-tuning of the cosmological constant to zero if all particles are massless. For massive particles one has to find a way to avoid the universal large cosmon-matter coupling $\beta=1/\sqrt{6}$, as we will discuss in the next section.

\section{$f(R)$-gravity with varying particle masses}
\label{$f(R)$-gravity with varying particle masses}

Having established the equivalence between $f(R)$-models and scalar-tensor theories a simple solution of the problem of a too large cosmon-matter coupling becomes visible. One may follow the strategy (i) in sect. \ref{Scalar tensor models}: If particle masses scale $~\sqrt{\phi}$ in the Jordan frame, their mass will be constant in the Einstein frame, implying $\beta=0$. Realistic models may therefore be found if the particle masses show an appropriate effective field dependence in the Jordan frame.

Let us consider the quarks and charged leptons. In the standard model of particle physics their masses are proportional to the expectation value $h_0$ of the Higgs doublet $h$. For cosmology, $h_0$ is replaced by the value of $h$ according to the cosmological solution. If this solution implies that $h_0$ scales proportional to $\sqrt{\phi}$ we will find a vanishing cosmon-matter coupling $\beta=0$ in the Einstein frame.

To be specific, we consider a first model where the effective action for gravity and the Higgs doublet is given by
\ba\label{VM1}
\Gamma&=&\int_x\sqrt{g}\left\{-\frac{a}{2}\left(\frac{R-2\mu^2}{2\epsilon}\right)^2 -h^{\dag}h\left(\frac{R-2\mu^2}{2\epsilon}\right)\right.\nn\\
&&\qquad\qquad+\frac{Z_h}{2}\partial^{\mu}h^{\dag}\partial_{\mu}h\left.\right\}.
\ea
The parameters $a$ and $\epsilon$ are dimensionless, such that scale symmetry is violated only by the parameter $\mu$ with dimension of mass. The function $f(y)$ is quadratic in $y$, with field dependent coefficient of the linear term,
\be\label{VM2}
f=a\left(\frac{y-2\mu^2/M^2}{2\epsilon}\right)^2+
\frac{2h^{\dag}h}{M^2}\left(\frac{y-2\mu^2/M^2}{2\epsilon}\right).
\ee
We emphasize that the Planck mass $M$ is not a parameter of the model \eqref{VM1}. In eq. \eqref{VM2} it is merely introduced by the conventions for $y$ and $f$.

According to eq. \eqref{GF4} the relation between $\phi$ and $R$ reads
\be\label{VM3}
\phi=\frac{a}{4\epsilon^2}(R-2\mu^2)+\frac{h^\dagger h}{2\epsilon},
\ee
and the corresponding potential of the equivalent scalar-tensor model becomes
\be\label{VM4}
V=\frac{1}{2a}(h^\dagger h-2\epsilon\phi)^2+2\mu^2\phi.
\ee
Identifying $2\phi=\chi^2$ we can associate the first term in eq. \eqref{VM4} with eq. \eqref{10}, for $\epsilon_h\sim \epsilon$ and $\lambda_h\sim 1/a$. For $h=h_0$ the potential becomes $V=2\mu^2\phi=\mu^2\chi^2$, which coincides for large $\chi$ with the potential \eqref{SF3}. 

In the Einstein frame the Higgs doublet is rescaled according to
\be\label{VM6}
h'=wh~,~w^2=\frac{M^2}{2\phi}.
\ee
This yields for the potential
\be\label{VM7}
V'=\frac{1}{2a}(h^{'\dagger}h'-\epsilon M^2)^2+\frac{\mu^2 M^4}{2\phi}.
\ee
It is obvious that $h'$ settles to a constant value at the minimum of $V'$, implying constant particle masses if the dimensionless Yukawa couplings are constant, $\beta=0$. 

The kinetic terms for $h'$ and $\phi$ in the Einstein frame read
\ba\label{VM8}
{\cal L}_{\rm kin}&=&\frac{Z_h}{2}
\big\{\partial^\mu h^{'\dagger}\partial_\mu h'+\frac12\partial^\mu\ln \phi~ \partial_\mu (h^{'\dagger}h')\big\}\nn\\
&&+\frac18(6M^2+Z_hh^{'\dagger}h')\partial^\mu\ln \phi~\partial_\mu\ln \phi.
\ea
For constant $h^{'\dagger}h'=\epsilon M^2$ the remaining kinetic term for $\phi$ becomes
\be\label{VM9}
{\cal L}_{\rm kin}=\frac{M^2}{8}
(6+Z_h\epsilon)\partial^\mu\ln \phi~\partial_\mu\ln \phi.
\ee
Neglecting the contribution $\sim Z_h\epsilon$ (see below) the normalized scalar field is related to $\phi$ by eq. \eqref{EX6} and eq. \eqref{36} remains valid. For $h'=h'_0$ the potential decays exponentially
\be\label{VM10}
V'=\mu^2 M^2\exp 
\left(-\frac{\alpha\varphi}{M}\right)~,~\alpha=\sqrt{\frac23}.
\ee

The value of $\alpha$ is too small for allowing for the scaling solutions with constant early dark energy fraction $\Omega_e<1$. This issue is related to the absence of a kinetic term for $\phi$ in eq. \eqref{46} or \eqref{GF1}. For initial values of $\phi_{\rm in}$ much smaller than $M^2$ the universe becomes scalar dominated long before the present epoch, leading to unrealistic cosmology. For $\phi_{\rm in}\gg M^2$ the scalar potential will play a role only in the far future and the model cannot account for dark energy. Realistic cosmology requires a particular initial value with $\phi_{\rm in}$ close to $M^2/2$. Cosmology is then of the type of ``thawing quintessence''. The need for a particular choice of initial conditions makes the model perhaps less attractive than the scaling solution found in the model of sect. \ref{Slow Freeze Universe}. 

Despite this shortcoming, the simple model \eqref{VM1} offers an interesting perspective on a dynamical fine tuning of the cosmological constant. Indeed, the effective cosmological constant vanishes asymptotically in the Einstein frame, even if we add an additional constant to the modified gravitational action \eqref{VM1}. In the Einstein frame the resulting contribution to $V'(\varphi)$ decays exponentially for large $\varphi$.  Scale symmetry becomes exact for $\varphi\to\infty$ and the cosmon corresponds in this limit to the dilaton, the Goldstone boson associated to the spontaneous breaking of scale symmetry. 

It is also interesting to discuss the issue of dilatation symmetry in the framework of $f(R)$-models. For $\mu=0$ the effective action \eqref{VM1} is scale invariant. The potential in the Einstein frame \eqref{VM4} has then one exactly massless direction, realizing the Goldstone boson. This demonstrates how the expected Goldstone boson arises in a model \eqref{VM1} that does not contain an explicit scalar singlet degree of freedom. 

The model \eqref{VM1} contains large dimensionless parameters. The Fermi scale is given by the canonically normalized doublet in the Einstein frame, $h_R=Z^{1/2}_hh'_0=175$GeV. This implies $\epsilon_h=Z_h\epsilon=(h_R/M)^2\approx 5\cdot 10^{-33}$. The renormalized quartic Higgs coupling is $\lambda_h=1/(aZ^2_h)$, such that the prefactor of $R^2$ in eq. \eqref{VM1} becomes $a/(8\epsilon^2)=1/(8\lambda_h\epsilon^2_h)\approx 10^{64}/(2\lambda_h)$, similar in size to eq. \eqref{F3a}. 

More reasonable couplings arise if one associates $h$ with a scalar field in some grand unified theory instead of the Higgs doublet. In this event $\epsilon_h$ could be roughly of the order one. The effective quark and lepton masses are then suppressed by the gauge hierarchy, i.e. the ratio between the Fermi scale and the scale $h_0$ which is now characteristic for grand unification. If gauge couplings take a fixed 
value for momenta given by $h$ also the QCD-confinement scale and therefore the nucleon masses are proportional to $h$, completing our mechanism for vanishing $\beta$. If $h$ is associated with a field characteristic for grand unification the parameter $a/\epsilon^2$ can be taken to be of the order one, such that the prefactor of the term $\sim R^2$ in eq. \eqref{VM1} is of the order one. In this case, however, $\phi$ is given essentially by $h^\dagger h/(2\epsilon)$ and the term $\sim R^2$ in eq. \eqref{VM1} plays only a negligible role. (The limit $a\to 0$ has no qualitative influence on the late cosmology of this model.)

As a second example we consider a family of models 
\be\label{N1}
\Gamma=\int_x\sqrt{g}
\left\{ 
\sigma(R^2+\rho)^{\frac\gamma 2}+\bar\lambda_c-\frac{1}{2\epsilon}h^\dagger hR+\frac{Z_h}{2}\partial^\mu h^\dagger\partial_\mu h\right\}.
\ee
The relation between $\phi,h$ and $R$ reads
\be\label{N2}
x=-\gamma\tilde\sigma y(y^2+\tilde \rho)^{\frac\gamma 2-1},
\ee
with 
\ba\label{N3}
x=\frac{2\epsilon\phi-h^\dagger h}{2M^2\epsilon}~,~y=\frac{R}{M^2}~,~\tilde\sigma=\sigma M^{2\gamma-4}~,~\tilde\rho
=\frac{\rho}{M^4}.\nn\\
\ea
In terms of $\phi$ the effective action becomes
\be\label{N4}
\Gamma=\int_x\sqrt{g}
\left\{ -\phi R+V(\phi,h)+\frac{Z_h}{2}\partial^\mu h^\dagger \partial_\mu h\right\},
\ee
where 
\be\label{N5}
V=M^4\tilde \sigma (y^2+\tilde\rho)^{\frac\gamma 2-1}
\left\{\tilde \rho +(1-\gamma)y^2\right\}+\bar\lambda_c,
\ee
and $y$ is related to $\phi$ and $h$ by eq. \eqref{N2}. After Weyl scaling the effective action for the metric and the scalars $\phi$ and $h'$ takes a standard form
\be\label{N6}
\Gamma=\int_x\sqrt{g'}
\left\{-\frac{M^2}{2}R'+V'(\phi,h')+{\cal L}_{\rm kin}\right\},
\ee
with 
\be\label{N7}
V'(\phi,h')=\frac{M^4 V}{4\phi^2},
\ee
and ${\cal L}_{\rm kin}$ given by eq. \eqref{VM8}. Again, $y$ is related to $x$ by eq. \eqref{N2} with
\be\label{N8}
x=\frac{\phi(\epsilon M^2-h^{'\dagger}h')}{\epsilon M^4}.
\ee

We may next investigate the field equation for $h'$. One finds a static solution with $h^{'\dagger}_0h'_0=\epsilon M^2$ provided $V(x)$ has its minimum for $x=0$. Particle masses are then constant in the Einstein frame, $\beta=0$. Inserting $h^{'\dagger}h'=\epsilon M^2$ and assuming $y(x=0)=0$ the potential gets a simple form
\be\label{N9}
V'=\frac{M^4}{4\phi^2}\left(\bar\lambda_c+\sigma \rho^{\frac\gamma 2}\right).
\ee
For positive $V_0=\bar\lambda_c+\sigma\rho^{\gamma/2}$ it decays to zero for $\phi\to \infty$. For a canonical scalar field (neglecting the term $Z_h\epsilon$ in eq. \eqref{VM9}) the potential decays exponentially
\be\label{N10}
V'=V_0\exp 
\left(-\frac{\alpha\varphi}{M}\right)~,~\alpha=\sqrt{\frac83}.
\ee
Again, this value of $\alpha$ is too small in order to realize the scaling solution with $\Omega_e<1$. Cosmology is similar to our first example, with realistic thawing quintessence realized for initial values $\phi_{\rm in}$ close to $10^{60}\sqrt{V_0}$. 

We notice that cosmology is the same for all ranges of $\gamma,\sigma$ and $\rho$ for which $V$ has its minimum for $x=0$. For $\rho>0$ the effective action \eqref{N1} and the potential $V$ are analytic. A special case occurs for $\rho=0$ which is similar to the model \eqref{EX1} except for the additional coupling to $h$. The potential is no longer analytic
\ba\label{N11}
V&=&M^4\tilde\sigma(1-\gamma)|y|^\gamma+\bar\lambda_c\nn\\
&=&M^4\tilde\sigma(1-\gamma)\left|\frac{x}{\gamma\tilde\sigma}\right|^{\frac{\gamma}{\gamma-1}}+\bar\lambda_c.
\ea
For $\rho>0$ the potential \eqref{N11} describes the behavior for large $y^2\gg\tilde \rho$. We observe that for $\gamma<1$ the limit $x\to 0$ can be reached for $|y|\to 0$ or $|y|\to\infty$. If the potential minimum corresponds to the second case the value $V_0=\bar\lambda_c$ may only be reached for asymptotic time $t\to\infty$.

We conclude that the problematic universal cosmon-matter coupling $\beta$ in the Einstein frame can be avoided if $f(R)$-theories allow for a suitable field dependence of particle masses. The other generic problem of $f(R)$-models, namely the need of large couplings multiplying the terms in a Taylor expansion of $f(y)$, will need a particular physics explanation which produces and stabilizes such large couplings appearing in the effective action. (In the generic case quantum fluctuations lead to a very fast running of very large dimensionless couplings, typically bringing them to values of the order one or making them divergent.) At present, we are still far from constructing an $f(R)$-model which would show a similar simplicity as the scalar-tensor theory discussed in sect. \ref{Slow Freeze Universe}. The benefit would be, of course, that no explicit scalar field $\chi$ is needed in modified gravity. 

\section{Non-local gravity}
\label{Non-local gravity}

For non-local gravity (see ref. \cite{Wo} for a recent review and references) the action involves the inverse of the covariant Laplacian ${\cal D}$, or similar operators that grow strongly for small covariant momenta. As a consequence, such modifications of gravity can play a role at long distances, without invoking very large dimensionless parameters as $\alpha$ in the preceding section. Already the first non-local gravity model in this spirit \cite{CWNL} has noted the equivalence to a model of a scalar field coupled to gravity.

Let us consider the effective action \cite{CWNL}
\be\label{NL1}
{\cal L}_g=\frac{M^2}{2}
\left\{-R+\frac{\tau^2}{2}R{\cal D}^{-1}R\right\},
\ee
with covariant derivative $D_\mu$ and covariant Laplacian
\be\label{NL2}
{\cal D}=-{D}^\mu D_\mu.
\ee
(In order to make eq. \eqref{NL1} well defined one has to regularize the operator ${\cal D}^{-1}$ \cite{CWNL}.) The model \eqref{NL1} admits an equivalent formulation as a scalar-tensor model with effective action
\be\label{NL3}
\Gamma=\int_x\sqrt{g}
\left\{-\frac{M^2}{2}(1+\tau\phi)R-\frac{M^2}{4}\partial^\mu\phi\partial_\mu\phi\right\}.
\ee
Indeed, the field equation for $\phi$, 
\be\label{NL4}
{\cal D}\phi=-D^\mu D_\mu\phi=-\tau R,
\ee
expresses $\phi$ as a functional of the metric,
\be\label{NL5}
\phi=-\tau {\cal D}^{-1}R.
\ee
Inserting the formal solution \eqref{NL5} into the action \eqref{NL3} yields the equivalent effective action \eqref{NL1} of non-local gravity.

The scalar-tensor theory \eqref{NL3} can be brought to the standard form of a coupled quintessence model by use of a Weyl scaling with 
\be\label{NL6}
w=(1+\tau\phi)^{-\frac12}.
\ee
The resulting kinetic term,
\be\label{NL7}
{\cal L}_{\rm kin}=\frac{M^2}{4}
\left(\frac{3\tau^2}{(1+\tau\phi)^2}-\frac{1}{1+\tau\phi}\right)
\partial^\mu\phi\partial_\mu\phi,
\ee
can be cast into a standard normalization \eqref{7} by defining $\varphi$ with
\be\label{NL8}
\frac{\partial\varphi}{\partial\phi}=\frac{M}{\sqrt{2}(1+\tau\phi)}
\sqrt{3\tau^2-(1+\tau\phi)}.
\ee
The potential vanishes for this model, similar to Brans-Dicke theory. 

The Weyl scaling typically leads to coupled quintessence. Consider non-local modified gravity \eqref{NL1} and a particle with constant mass $m$. One obtains in the Einstein frame a $\varphi$-dependent mass, $m'=w(\varphi)m$. Defining the $\varphi$-dependent coupling $\beta(\varphi)$ by 
\be\label{NL9}
\beta(\varphi)=-M
\frac{\partial\ln m'}{\partial\varphi}
\ee
one obtains
\be\label{NL10}
\beta=\left[6-\frac{2}{\tau}\left (\phi+\frac1\tau\right)\right]^{-\frac12},
\ee
where $\phi$ can be expressed in terms of $\varphi$ using eq. \eqref{NL8}. 

We observe that stability requires a positive effective Planck mass and a positive kinetic term \eqref{NL7}, which is realized for the range
\be\label{NL11}
0\leq 1+\tau\phi\leq 3\tau^2.
\ee
In this range $\beta$ is well defined. The minimum value for $\beta$ is
\be\label{NL12}
\beta_{\rm min}=\frac{1}{\sqrt{6}},
\ee
resembling a Brans-Dicke theory with $\omega=0$. Such a large coupling is not compatible with observation, such that the model \eqref{NL1} is not phenomenologically viable \cite{CWNL}.

In summary, the gravitational part of non-local gravity models has no problem of consistency. It is equivalent to standard gravity coupled to a massless scalar, similar to Brans-Dicke theory. Adding relativistic particles as photons remains unproblematic. Issues of compatibility with observation arise, however, if massive particles are considered within non-local gravity. The coupling between the scalar field and massive particles typically turns out to be unacceptably large.

One may construct large classes of consistent non-local gravity models by starting from a local scalar-tensor model that only contains terms linear and quadratic in $\phi$. Such generalizations of eq. \eqref{NL3} can contain higher derivatives of $\phi$, a coupling of $\phi$ to higher order curvature invariants, terms $\sim R\partial^\mu\phi\partial_\mu\phi$ or $\sim R^{\mu\nu}\partial_\mu\phi\partial_\nu\phi$ etc.. The field equations for $\phi$ involve terms linear in $\phi$ as well as a $\phi$-independent ``source term''. The general solutions are functionals of the metric. Inserting these solutions into the action yields consistent models of non-local gravity. Consistency does not imply compatibility with observation, however. It seems not easy to avoid a too large coupling between the scalar field and massive particles in the Einstein frame. 

While non-local modifications of gravity are consistent, it is not easy to motivate why the quantum effective action for gravity should have this form. Unless one can identify some quantum effect producing such non-localities they may not look very natural, however. For the moment, the only physically well motivated origin of non-localities of the type discussed in this section that is known to us arises from the exchange of an effective massless degree of freedom, similar to the Coulomb interaction between electrons or the Newtonian interaction between massive particles. In this event it seems much simpler to use directly a field for the exchanged particle.

\section{Higher derivative modified gravity with second order field equations}
\label{Higher derivative}

We have seen that $f(R)$-theories and a large class of non-local gravity theories can be mapped to a quintessence model,
\be\label{c1}
\Gamma=\int_x\sqrt{g'}
\left\{-\frac12 M^2R'+\frac12\partial^\mu\varphi\partial_\mu\varphi+V(\varphi)\right\},
\ee
by an appropriate Weyl scaling. One may ask how large is the class of modified gravity theories that can be mapped to the simple action \eqref{c1} by suitable field transformations. A large class of actions involving higher derivatives, that nevertheless lead to second order field equations, has been found by Horndeski \cite{Horn}. One would like to know if they are equivalent to the action \eqref{c1}. 

Part of the answer can be given by considering general field transformations
\ba\label{c2}
\varphi&=&v(\chi,R,\partial^\mu\chi\partial_\mu\chi,\dots)\nn\\
g'_{\mu\nu}&=&w^{-2}(\chi,R,\partial^\mu\chi\partial_\mu\chi,\dots)g_{\mu\nu}\nn\\
&&+s_1(\chi,R,\partial^\mu\chi\partial_\mu\chi,\dots)\partial_\mu\chi\partial_\nu\chi\nn\\
&&+s_2(\chi,R,\partial^\mu\chi\partial_\mu\chi,\dots)R_{\mu\nu}+\dots
\ea
Here $v,w,s_1,s_2$ are functions of various possible scalars that can be formed from $\chi$ and $g_{\mu\nu}{'}$, with dots standing for additional scalars as $R_{\mu\nu}R^{\mu\nu}$, $\partial_\mu\chi\partial_\nu\chi R^{\mu\nu}$ etc.. We only require that the objects on the r.h.s. of eq. \eqref{c2} have the correct tensor transformation properties. 

It is obvious that a very large class of effective actions for modified gravity can be constructed by inserting eq. \eqref{c2} into eq. \eqref{c1}. 
\be\label{c2A}
\Gamma[\chi,g_{\mu\nu}]=\Gamma\big[\varphi[\chi,g_{\mu\nu}]~,~g'_{\rho\sigma}[\varphi,g_{\mu\nu}]\big].
\ee
All these models have as physical degrees of freedom a scalar coupled to the graviton. Even though these actions can contain an arbitrary number of derivatives, the field equations will finally be second order field equations, equivalent to those derived from the action \eqref{c1}. The requirement of equivalence imposes, however, some mild conditions on the functions appearing in eq. \eqref{c2}. What is needed is the invertibility of the variable transformation \eqref{c2}.

We may demonstrate this explicitly for transformations with $s_1=s_2=0$. The field equations for the transformed action, 
\be\label{c3}
\frac{\partial\Gamma}{\partial\chi(x)}=0~,~\frac{\partial\Gamma}{\partial g_{\mu\nu}(x)}=0,
\ee
can be expressed as ($\partial$ stands here for functional derivatives)
\be\label{c4}
\int_y
\left\{\frac{\partial\Gamma}{\partial g'_{\mu\nu}(y)}
\frac{\partial w^{-2}(y)}{\partial\chi(x)}
g_{\mu\nu}(y)+
\frac{\partial\Gamma}{\partial\varphi(y)}
\frac{\partial v(y)}{\partial\chi(x)}\right\}=0,
\ee
and 
\ba\label{c5}
&&\int_y
\left\{\frac{\partial\Gamma}{\partial g'_{\rho\sigma}(y)}
\frac{\partial w^{-2}(y)}{\partial g_{\mu\nu}(x)}
g_{\rho\sigma}(y)+w^{-2}(y)
\frac{\partial\Gamma}{\partial g'_{\mu\nu}(x)}
\delta(y-x)\right.\nn\\
&&\hspace{3.0cm} \left.+\frac{\partial\Gamma}{\partial\varphi(y)}
\frac{\partial v(y)}{\partial g_{\mu\nu}(x)}\right\}=0.
\ea
Obviously, the solutions of the field equations of the action \eqref{c1},
\be\label{c6}
\frac{\partial\Gamma}{\partial \varphi(y)}=0~,~\frac{\partial\Gamma}{\partial g'_{\mu\nu}(y)}=0,
\ee
are also solutions of the field equations \eqref{c3}. The conditions on the functions $w$ and $v$ have to ensure that no additional ``spurious'' solutions are generated by the transformation \eqref{c2}. 

Consider, for example, the case $w=1$. Then the matrix $\partial v(y)/\partial\chi(x)$ should be invertible, such that eq. \eqref{c4} implies $\partial\Gamma/\partial\varphi(y)=0$. Invertibility means that a function $H(x,z)$ exists such that 
\be\label{c7}
\int_x\frac{\partial v(y)}{\partial\chi(x)}H(x,z)=\delta(y-z).
\ee
For $w=1$ the gravitational field equation \eqref{c5} reads
\be\label{c8}
\frac{\partial\Gamma}{\partial g'_{\mu\nu}(x)}+\int_y
\left\{\frac{\partial\Gamma}{\partial\varphi(y)}\frac{\partial v(y)}{\partial g_{\mu\nu}(x)}\right\}=0.
\ee
The second term vanishes for invertible $\partial v(y)/\partial\chi(x)$ since $\delta\Gamma/\delta\varphi(y)=0$, such that both field equations \eqref{c6} must be obeyed necessarily. Similarly, we may consider $v=\chi$ and an invertible matrix $\partial g'_{\rho\sigma}(y)/\partial g_{\mu\nu}(x)$. The field equation \eqref{c5} implies then $\partial\Gamma/\partial g'_{\mu\nu}(x)=0$, such that eq. \eqref{c4} guarantees $\partial\Gamma/\partial\varphi(x)=0$. Again, the field equations \eqref{c6} must be necessarily obeyed. This generalizes to arbitrary transformations $g_{\rho\sigma}(y)\big[g_{\mu\nu}(x)\big]$, as in eq. \eqref{c2}. Invertible transformations with $v=\chi$ or $w=1$ can be combined to yield more general invertible transformations. We conclude that invertibility of the transformation \eqref{c2} guarantees the absence of spurious solutions, such that the effective action $\Gamma[g_{\mu\nu},\chi]$ is fully equivalent to $\Gamma[g'_{\mu\nu},\varphi]$ given by eq. \eqref{c1}.

It may be instructive to discuss two simple examples of field transformations with $w=1$. For the first we take $\varphi=v(\chi,R)$, such that 
\be\label{c9}
\frac{\partial v(y)}{\partial\chi(x)}=\frac{\partial v}{\partial\chi}\big(\chi(x),R(x)\big)\delta(y-x).
\ee
If $\partial v/\partial \chi$ is non-vanishing for all $\chi$ and $R$ the transformation is invertible. On the other hand, if $\partial v/\partial\chi=0$ has a solution $\chi_0(R)$, the configuration $\chi=\chi_0(R)$ solves the field equation $\partial\Gamma/\partial\chi(x)=0$ without being a solution of eq. \eqref{c6}. This is an example of a spurious solution. A second example with a spurious solution is 
\be\label{c10}
\varphi(x)=m^{-3}\big(\chi;^\mu{_\mu}(x)+m^2\chi(x)\big)\chi(x).
\ee
While the solutions \eqref{c6} remain solutions of the field equations \eqref{c3}, additional solutions of eq. \eqref{c3} are provided by $\chi;^\mu_\mu+m^2\chi=0$. This model can still be cast into the form of an action with at most two derivatives, involving two scalar fields. Besides the solutions \eqref{c6} one has new solutions for non-zero values of a free massive scalar field with mass $m$. (The last term in eq. \eqref{c5} ensures that the energy momentum tensor of the second scalar field is induced in the gravitational field equation.) Many transformations with higher derivatives are invertible and do not lead to spurious solutions, however.

It remains an interesting question if invertible transformations of the type \eqref{c2} are sufficient in order to show the equivalence of a large class (or all) of Horndeski's models with the effective action \eqref{c1}. This seems very likely to us for models that contain no further physical degrees of freedom besides a scalar and the graviton. The effective action \eqref{c2A} obtained by inserting eq. \eqref{c2} into eq. \eqref{c1} may even lead to still larger classes of higher derivative modified gravity for which all cosmological solutions can be obtained from second order field equations. Further generalizations are possible if one adds scalar, vector or tensor fields with no more than two derivatives to eq. \eqref{c1}, and subsequently makes a field transformation of the type \eqref{c2}. 

The field transformations \eqref{c2} are a convenient way to construct effective actions \eqref{c2A} that only involve second order field equations for the scalar-graviton system. This does not mean that all models based on an action \eqref{c2A} are equivalent to those based on the action \eqref{c1}. The field transformations also affect the matter part ${\cal L}_m$. Consider a model where matter is minimally coupled to the metric $g_{\mu\nu}$ and particle masses are $\chi$-independent. It becomes typically a model of coupled quintessence with non-minimal gravitational interactions once written in terms of $g'_{\mu\nu}$ and $\varphi$. The inverse of the transformation \eqref{c2}, which maps the action \eqref{c2A} onto \eqref{c1}, can induce in the matter and radiation sector a complicated dependence on $\varphi$ and $g'_{\mu\nu}$. Even if we approximate ${\cal L}_m$ in the generalized Jordan frame \eqref{c2A} by free massive or massless particles, non-trivial interactions will appear in the Einstein frame \
eqref{c1}. This is the way how the functions $v,w,s_1,s_2$ in eq. \eqref{c2} can affect the predictions for observations. Similar to $f(R)$-models also, the much more general class of models \eqref{c2A} encounters often problems with too large effective couplings $\sim\beta$ in the Einstein frame.

\section{Conclusions}
\label{Conclusions}

Can one distinguish modified gravity from dark energy by observation? In view of the equivalence of a large class of modified gravity models with coupled quintessence an answer to this question is not straightforward. Statements that modified gravity and quintessence lead to different growth factors for cosmic structures apply only to quintessence models without coupling to matter. We have seen, however, that the quintessence models that are equivalent to modified gravity typically have a nonzero coupling $\beta$ between the cosmon and different forms of matter. (This coupling needs not to be the same for all species of massive particles.) In this view precision measurements of the growth rate can differentiate between uncoupled and coupled quintessence and determine bounds on $\beta$. The issue if there are modified gravity models that can be distinguished observationally from coupled quintessence is much harder to answer.

Modified gravity models almost always involve new degrees of freedom besides the graviton. This is a consequence of the fact that models for a massless spin two particle are severely constrained by consistency requirements. The conjecture that consistency requires diffeomorphism symmetry (more precisely its unimodular subgroup) has never been proven, but no counter examples are known either. A model containing a massless spin two particle as the only degree of freedom is then rather close to general relativity. Modifications of gravity therefore typically involve additional degrees of freedom, as scalars or massive spin two particles. 

The field description of the additional degrees of freedom is not unique. For example, a scalar may be described as a component of the metric (modified gravity) or by a separate field (quintessence). Very large classes of models can be mapped onto each other by non-linear field transformations. Field relativity states that observables cannot depend on the choice of fields. For models related by field transformations no observational distinction is possible. We have seen that this holds for variable gravity models where the Planck mass is field dependent. It also applies to $f(R)$-models and large classes of non-local gravity. Very general models equivalent to coupled quintessence models have been discussed in the preceding section. 

For all these models modified gravity and coupled quintessence should merely be seen as two different pictures describing the same reality, in analogy to the Jordan frame and Einstein frame for the metric. For practical computations of the evolution of homogeneous cosmology and fluctuations around this background the simplest way uses the Einstein frame. This holds both for the linear treatment of fluctuations and for numerical simulations in the non-linear regime. The physical effects of the cosmon-matter coupling $\beta$ are intuitively accessible in the Einstein frame. 

For modified gravity models that are equivalent to coupled quintessence one may ask: why then discuss them all? If there is no observational distinction, the discussion of such modifications of gravity may at first sight look like a redundant exercise. A deeper answer concerns questions of simplicity and naturalness. Models of modified gravity can be very simple and involve no unnatural parameters. Nevertheless, the equivalent description in the Einstein frame by coupled quintessence may hide simplicity and naturalness in the complexity of the field transformation. An example is the big bang singularity. We have presented in sect. \ref{Slow Freeze Universe} a modified gravity model for which the ``beginning'' of the universe is very slow and cold. It has no big bang singularity, the cosmological solution can be continued to the infinite past $t\to -\infty$. In the Einstein frame the same model is described as a hot big bang. Models may be regular in the Jordan frame and show a big bang singularity in 
the Einstein frame. This singularity is then due to a singularity in the field transformation  \cite{CWU}, in close analogy to a coordinate singularity.

The question of naturalness is often closely linked to symmetries. Scale symmetry is explicitly visible in the modified gravity description of the models in sects. \ref{Scalar tensor models}, \ref{Slow Freeze Universe}. It is realized by a multiplicative rescaling of the metric and the scalar field $\chi$. In the presence of quantum fluctuations scale symmetry is violated by $\chi$-dependent (``running'') dimensionless couplings. For fixed points of the running exact (quantum-) scale symmetry is restored. For the quantum effective action \eqref{SF2} such fixed points are present for $\chi\to 0$ and $\chi\to\infty$ \cite{Wetterich:2014eaa}. 

In our model in sect. \ref{Slow Freeze Universe} the asymptotic value 
\be\label{CCA}
\lambda_\infty=\lim_{\chi\to\infty} V(\chi)/\chi^4
\ee
vanishes for the fixed point at $\chi\to\infty$. This can be motivated by properties of a possible ultraviolet fixed point in dilaton quantum gravity \cite{RP} or by dilatation symmetry in higher dimensions \cite{CWDL1,CWDL2}. The fixed point with $\lambda_\infty=0$ is the deeper reason why the cosmological constant vanishes asymptotically in the Einstein frame, $\lim_{\varphi\to\infty}V'(\varphi)\to 0$. Without this understanding of naturalness as a consequence of fixed point properties one would argue in the Einstein frame that naturalness suggests the addition of a constant to eq. \eqref{F6}. Apparently convincing qualitative arguments on the induction of a cosmological constant by quantum fluctuations in the Einstein frame yield very different results when applied in the Jordan frame. A constant term in $V(\chi)$ yields a term $V'(\varphi)\sim \exp (-2\alpha\varphi/M)$ in the Einstein frame which vanishes for $\varphi\to\infty$. This is one more example how modified gravity can shed new light on 
questions of 
naturalness. 

The possibility of field transformations from modified gravity theories to coupled quintessence models in the Einstein frame is an extremely useful tool for the discussion of observational consequences of a model. It should not prevent us, however, to look for modified gravity theories distinguished by simplicity and naturalness.

\newpage

\bibliography{draft_modified_gravity}

\end{document}